\documentclass[11pt]{article}
\usepackage{amsfonts}

\newcommand{\sect}[1]{\setcounter{equation}{0}\section{#1}}

\renewcommand{\theequation}{\arabic{section}.\arabic{equation}}

\def\be{\begin{equation}}
\def\ee{\end{equation}}
\def\bea{\begin{eqnarray}}
\def\eea{\end{eqnarray}}

\def\R{{\mathbb R}}
\def\1{\'{\i}}                           

\def\gal{{\cal G}}
\def\poinc{{\cal P}}

\def\jja{{\cal J}_1}
\def\ppa{{\cal C}_1}
\def\jjb{{\cal J}_2}
\def\ppb{{\cal C}_2}
\def\jjj{{\cal J}}

\def\parity{\Pi}
\def\timereversal{{\rm T}}

\def\expandcasa{{\cal C}'_1}
\def\expandcasb{{\cal C}'_2}

\def\ext{m}
\def\AA{\alpha}
\def\ww#1{W_{#1}}

\def\>#1{{\bf #1}}

\begin{document}
 
\thispagestyle{empty}

\  
\hfill  math-ph/9909005

\vspace{2cm}

\begin{center}
{\LARGE{\bf{A new Lie algebra expansion method:}}}
 
{\LARGE{\bf{Galilei expansions to Poincar\'e and Newton--Hooke}}}
\end{center}

\bigskip

\bigskip

\begin{center}
Francisco J. Herranz$^\dagger$ 
and Mariano Santander$^\ddagger$
\end{center}

\begin{center}
{\it $^\dagger$ Departamento de F\'{\i}sica, Escuela
Polit\'ecnica Superior\\ 
Universidad de Burgos, E--09006 Burgos, Spain}
\end{center}

\begin{center}
{\it $^{\ddagger}$ Departamento de F\'{\i}sica Te\'orica,
Facultad de Ciencias\\ Universidad de Valladolid,
E--47011 Valladolid, Spain}
\end{center}

\bigskip\bigskip\bigskip

\begin{abstract} 
We modify a Lie algebra expansion method recently  introduced for
the $(2+1)$-dimensional kinematical algebras so as to work for
higher dimensions. This new improved and geometrical procedure is
applied to expanding the $(3+1)$-dimensional Galilei algebra and
leads to its  physically meaningful `expanded' neighbours. One
expansion gives rise to the Poincar\'e algebra, introducing a
curvature $-1/c^2$ in the flat Galilean space of worldlines, while
keeping a flat spacetime which changes from absolute  to
relative time in the process. This formally reverses, at a Lie
algebra level, the well known non-relativistic contraction
$c\to \infty$ that goes from the Poincar\'e group to the Galilei
one; this expansion is done in an explicit constructive way. The
other possible expansion leads to the Newton--Hooke algebras,  
endowing with a  non-zero spacetime curvature $\pm 1/\tau^2$ the
spacetime, while keeping a flat space of worldlines.
\end{abstract}

\newpage 

%%%%%%%%%%%%%%%%% Introduction %%%%%%%%%%%%%%%%%%%%%%%%%%%%

\sect{Introduction}

Expansions of Lie algebras can be considered as the opposite
processes of  Lie algebra contractions. In general, starting
from a Lie algebra a {\em contraction} gives rise to  another
more abelian algebra by making  some  structure constants
to vanish \cite{IW}--\cite{Weimar}, while an {\em expansion}
goes to another less abelian algebra `producing' some non-zero 
structure constants \cite{Gilmore}--\cite{expansion}. The idea
of contractions of Lie algebras and groups historically appeared 
in relation with the non-relativistic limit where the
relativistic constant $c$ (the speed of light) goes to
infinity; that limit brings relativistic mechanics
(Poincar\'e group) to classical mechanics (Galilei group).
In the framework of kinematical algebras \cite{BLL,Bacry}, the
scheme of contractions  is well known 
\cite{BLL}: starting from de Sitter algebras a sequence of
different contractions leads to Poincar\'e,
Newton--Hooke, Galilei, \dots, ending up at the last stage
in the so called Static algebra. From the viewpoint of graded
contraction theory 
\cite{MooPat, MonPat}, the $(3+1)$-dimensional case has been 
studied in \cite{Tolar}, and kinematical contractions in
arbitrary dimension have been obtained in \cite{solgen}.

Unlike the  Lie algebra contractions, the theory of expansions
has not  been so systematized. As a kind of formal `inverse' of
contractions its status is clear, but as yet there is no a
general constructive theory where explicit realizations of an
`expanded' Lie algebra can be given in terms of an `initial' one.
However some specific procedures, valid for certain
algebras and certain expansions, have been introduced. Expansions
from the inhomogeneous pseudo-orthogonal algebras
$iso(p,q)$ to the semisimple ones $so(p+1,q)$ with $(p+q=N)$ can
be found in  \cite{Gilmore,Rosen}; these contain  as particular
cases those expansions starting from the Euclidean  algebra and
leading to either the elliptic or hyperbolic ones, and also those
expansions from Poincar\'e to both de Sitter algebras:
\be
\begin{array}{lll}
\mbox{Euclidean expans.:}
&\  iso(N)\to so(N+1)  &\ iso(N)\to so(N,1)  \cr
\mbox{Poincar\'e expans.:}
&\  iso(N-1,1)\to so(N,1)  &\  iso(N-1,1)\to so(N-1,2)  
\end{array}
\label{intro}
\ee
A different method   \cite{WB}
enables to perform    expansions  from $t_{qp}(so(p)\oplus
so(q))\to so(p,q)$. Similar expansions for unitary algebras can
be found in the above references. 

Recently, another new expansion method  was proposed in
\cite{expansion}. Although formulated in algebraic terms, the ideas 
follow from geometrical considerations. The main trait is to control 
the expansion by a parameter which is the curvature of
some homogeneous spaces  associated to both the initial and
expanded Lie algebras; the procedure rests on the Casimir
operators associated to both the initial and
expanded Lie algebras.  This method  was applied to   the
expansions within the set of  $(2+1)$-dimensional kinematical
algebras \cite{BLL}. In this
 $(2+1)$D case, known expansions as (\ref{intro}) were recovered
and, furthermore, other new expansions were established; amongst
them,  we remark  several algebra expansions which starting
from  Galilei  give  rise to either Poincar\'e or
Newton--Hooke, and further expansions going from
Newton--Hooke to de Sitter.

 In this method the emphasis is put in the structure of
homogeneous spaces associated to the two Lie groups involved in the
contraction/expansion. Contraction would be associated to vanishing
curvature and/or degenerating the metric, while expansion would mean to
produce non-vanishing curvature or to make the metric non-degenerate. 

A natural and physically interesting frame to discuss these questions
is provided by homogeneous models of spacetimes (see e.g.\ 
\cite{trigo}), which are either relativistic or non-relativistic, and
have a spacetime curvature which can be either zero or
non-zero (this would correspond to a cosmological constant).   It seems
interesting to perform a mathematical study of expansion procedures which
would work in these
$(3+1)$D physically relevant cases. However, when followed literally
the method in
\cite{expansion} do not work in the  $(3+1)$D case.  

In this paper we set two objectives. First, in the same geometrical
vein, we propose a new expansion procedure which, in principle, 
can be applied to any Lie algebra in any dimension; the method in
\cite{expansion} can be seen as a particular instance of the new
expansion method.  The second and   main goal of this paper is  to
apply this procedure to the Galilei algebra in the proper
kinematical dimension, the $(3+1)$D case. This allows us to produce,
in an explicit constructive way, {\em two} expansions which are
physically relevant. One reverses the well known 
non-relativistic limit at a Lie algebra level:  we 
start from the $(3+1)$D Galilei algebra and we obtain the
Poincar\'e one within the universal enveloping algebra of the
former. The other reverses the ordinary zero-curvature limit in
the non-relativistic Newton--Hooke spacetimes, whose Lie algebras
are obtained within the universal enveloping algebras of the
centrally extended $(3+1)$D Galilei algebra. 

The paper is organized as follows: in the next section we
recall the  basics of Galilei algebra structure and its main
associated symmetric homogeneous spaces (spacetime and space of
worldlines) showing two natural ways of expansion to either
Poincar\'e or Newton--Hooke algebras. In  section 3 we
summarize the steps of the `improved' expansion method that we
propose. Its application to the expansions going from
Galilei to  either Poincar\'e or Newton--Hooke algebras are
explicitly  developed in the sections 4 and 5,
respectively. Finally, some remarks are pointed out
in the last section.

%%%%%%%%%%%%%%%%% Galilei %%%%%%%%%%%%%%%%%%%%%%%%%%%%%%%%%%

\sect{The Galilei  algebra and associated homogeneous spaces}

Let $H$, $P_i$, $K_i$ and $J_i$ ($i=1,2,3$) be the usual
generators of time translation, space translations, boosts and
spatial rotations, respectively. The Lie brackets of the
$(3+1)$D Galilei algebra, $\gal$, are given by
\be
\begin{array}{lll}
[J_i,J_j]=\varepsilon_{ijk}J_k& \qquad
[J_i,P_j]=\varepsilon_{ijk}P_k&
\qquad [J_i,K_j]=\varepsilon_{ijk}K_k\cr
[P_i,P_j]=0&\qquad [P_i,K_j]=0 &
\qquad [K_i,K_j]=0\cr
[H,P_i]=0&\qquad [H,K_i]=-P_i &\qquad [H,J_i]=0  
\end{array}
\label{aa}
\ee
where $i,j,k=1,2,3$ and $\varepsilon_{ijk}$ is the completely
skewsymmetric tensor with $\varepsilon_{123}=1$.
Hereafter  any generator or object with three  components is
 denoted as  $\>X=(X_1,X_2,X_3)$; its `square' and its `product' 
with other element, say
$\>Y=(Y_1,Y_2,Y_3)$, are
\be
\>X^2=X_1^2+X_2^2+X_3^2\qquad \>X\>Y=X_1Y_1+X_2Y_2+X_3Y_3 .
\label{ad}
\ee

The Galilei algebra has two Casimir invariants  which read
\cite{casimir}:
 \be 
 \ppa=\>P^2=P_1^2+P_2^2+P_3^2\qquad
\ppb=\>W^2= \ww1^2+ \ww2^2+\ww3^2
\label{ab}
\ee
where the components of $\>W$ are given by
\be
\ww1= P_3K_2 -P_2 K_3 \qquad
\ww2=  P_1 K_3-P_3K_1 \qquad
\ww3=  P_2K_1 -P_1 K_2.
\label{ac} 
\ee
The  second-order Casimir $\ppa$  (which is related with the
Killing--Cartan form) corresponds in the free kinematics of a
particle in the Galilean spacetime to the square of the linear
momentum (i.e., to the non-relativistic energy), while the 
fourth-order invariant 
$\ppb$  can be identified with the square of the angular
momentum.

We remark that  the Galilei algebra is isomorphic to
a twice inhomogeneous orthogonal algebra:
\be
\gal \equiv iiso(3)\equiv t_4\odot (t_3\odot so(3))  
\label{aac}
\ee
with the two abelian subalgebras  $t_4$, $t_3$ and the
orthogonal subalgebra $so(3)$ spanned by
\be
t_4=\langle H,\>P\rangle\qquad 
t_3=\langle  \>K\rangle\qquad 
 so(3)=\langle \>J\rangle .
\label{aad}
\ee

As any kinematical group, the Galilei group generated by
the Lie algebra $\gal$  has two symmetric homogeneous
spaces identified with the  spacetime and  space of (time-like)
wordlines. According to the two involutive automorphisms parity 
$\parity$ and the product 
$\parity\timereversal$ ($\timereversal$ is the time-reversal),
defined by \cite{BLL}
\be
\begin{array}{ll}
\parity\timereversal: \quad &(H,\>P,\>K,\>J)\to 
(-H,-\>P,\>K,\>J)\cr
\parity: \quad &(H,\>P,\>K,\>J)\to  (H,-\>P,-\>K,\>J)
\end{array} 
\label{aae}
\ee
 we find  two Cartan Lie algebra decompositions given by
\be
\begin{array}{llll}
\parity\timereversal: \quad &\gal=p^{(1)}\oplus h^{(1)}&\quad
p^{(1)}=\langle H,\>P\rangle &\quad
h^{(1)}=\langle  \>K,\>J\rangle \cr
\parity: \quad &\gal=p^{(2)}\oplus h^{(2)}&\quad
p^{(2)}=\langle \>P,\>K\rangle &\quad
h^{(2)}=\langle
H,\>J\rangle= \langle
H\rangle \oplus \langle \>J\rangle 
\end{array} 
\label{ae}
\ee
fulfilling
\be
[h^{(l)},h^{(l)}]\subset h^{(l)} \qquad
[h^{(l)},p^{(l)}]\subset p^{(l)} \qquad [p^{(l)},p^{(l)}]=0
\qquad l=1,2 .
\label{af}  
\ee 
Notice that both $h^{(l)}$ and $p^{(l)}$  are Lie subalgebras;
the latter is an abelian one. Consequently, the Galilei group
$G\equiv IISO(3)$ is  the motion group of the following 
symmetrical homogeneous spaces:
\be
\begin{array}{lll}
{\cal S}^{(1)}=G/H^{(1)}=IISO(3)/ISO(3)&\quad
\mbox{dim}\,({\cal S}^{(1)})=3+1  &\quad
\mbox{curv}\,({\cal S}^{(1)})=0\cr
 {\cal S}^{(2)}=G/H^{(2)}=IISO(3)/(\R\otimes
SO(3))&\quad
\mbox{dim}\,({\cal S}^{(2)})=3+3 &\quad
\mbox{curv}\,({\cal S}^{(2)})=0 .  
\end{array} 
\label{ag}
\ee
The Galilei subgroups   $H^{(1)}$, $H^{(2)}$  (whose Lie
algebras are $h^{(1)}$, $h^{(2)}$) are the isotopy subgroups
of an event  and a time-like line, respectively. Therefore  
${\cal S}^{(1)}$ is identified with the   $(3+1)$D Galilean
spacetime, while  ${\cal S}^{(2)}$ is the 6D space of
time-like lines in the Galilean spacetime  ${\cal S}^{(1)}$. Both
spaces are of zero curvature: 
Galilean spacetime is a {\em flat} universe with a degenerate
metric of {\em absolute time}, and the set of time-like lines is
a {\em flat} rank-two space with a degenerate metric where 
distance corresponds to  relative velocity.

By taking into account  the spaces (\ref{ag}),  two  possible
Galilei algebra expansions arise in a natural way. First,  if
we consider the space of worldlines   ${\cal S}^{(2)}$ we can
try to obtain a Lie algebra for another kinematics whose
corresponding space of time-like lines has a curvature
different from zero but keeping a flat spacetime. This
expansion allows us to reach the Poincar\'e algebra by
introducing in a suitable way a (negative) curvature in 
${\cal S}^{(2)}$ equal to $-1/c^2$.  In this sense, this
process is a {\em relativistic expansion}  as it gives rise to
the Minkowskian spacetime which is also a  {\em flat} universe
but of {\em relative time}. This is exactly the opposite
process to the well known non-relativistic limit or
contraction studied by In\"on\"u and Wigner \cite{IW}, Segal
\cite{Segal} and   Saletan \cite{Saletan}. This contraction is
also called speed-space contraction \cite{BLL} and it
corresponds to  the limit $c\to
\infty$ (i.e.\ $-1/c^2\to 0$) in the Poincar\'e algebra.

The second  possibility is to start with the Galilean
spacetime ${\cal S}^{(1)}$ and to reach a Lie algebra whose
associated spacetime has a {\em non-zero} curvature but keeps a
{\em flat} (rank-two) space of worldlines. In this way we obtain
the two Newton--Hooke algebras; these are {\em non-relativistic}
expansions and  provide  {\em curved} spacetimes whose
curvature is $\kappa=\pm 1/\tau^2$ (where $\tau$ is the universe
`radius' measured in time units) but are still of {\em
absolute time}. The opposite process is the so called  spacetime
contraction \cite{BLL} characterized by the limit $\tau\to \infty$
(i.e.\ $\kappa\to 0$)  in the  Newton--Hooke algebras

%%%%%%%%%%%%%%%%% expansion method %%%%%%%%%%%%%%%%%%%

\sect{An `improved' expansion method}

 In a previous paper \cite{expansion} we proposed an expansion
method, which worked in the kinematical $(2+1)$D, where the Lie
algebras are contractions of $so(4)$. This is a (only semisimple)
rank-two algebra. 
Furthermore, $so(n)$ has as many Casimirs as its rank, one quadratic
and the others higher order polynomials in the generators
with an exceptional behaviour in $so(4)$, where the additional Casimir
is also essentially quadratic. Then it could happen that a method
working for $so(4)$ may not be directly extensible to higher
dimensions. Thus this expansion method  should be replaced by a more
general one. A proposal, which keeps the geometric flavour of the
previous method can be described as follows.

Let $g$ a Lie algebra which is obtained as a contraction from
another Lie algebra $g'$: $g'\to g$. Suppose that the
contraction corresponds to making equal to zero the {\em
curvature} $\omega$ of some homogeneous spaces associated to $g$
and $g'$ (by taking the quotient associated to  the common
subalgebra invariant under the contraction). Assume for
simplicity that $g'$, $g$ are rank-two algebras and let
${\cal C}_1$, ${\cal C}_2$   the two Casimirs of the  initial Lie
algebra $g$ (with $\omega=0$) and ${\cal C}'_1$, ${\cal C}'_2$
those of the final algebra $g'$ (with $\omega\neq 0$). The main
steps of our expansion method are:

\noindent
(i) Write each expanded Casimir as polynomials on the 
curvature $\omega$ we aim to recover:
\be
{\cal C}'_1 = {\cal C}_1 +\omega {\cal J}_1 +
\omega^2 {\cal M}_1+\dots \qquad
{\cal C}'_2 = {\cal C}_2 +\omega {\cal J}_2  + \omega^2 {\cal
M}_2+\dots .
\label{bbba}
\ee
where ${\cal
C}_l$, ${\cal J}_l$,  ${\cal M}_l$,\dots ($l=1,2$) are {\em
independent} of $\omega$. Obviously the terms which are
zero-order in
$\omega$ are just the `contracted' Casimirs ${\cal C}_l$.

\noindent
(ii) Assume to work  in the universal enveloping 
algebra of the initial Lie algebra $g$ within an irreducible
representation and  consider as  `expansion seed' a linear
combination formed by the terms {\em linear} in the curvature
$\omega$:
\be
{\cal J}=\alpha_1 {\cal J}_1 + \alpha_2{\cal J}_2 
\label{bbbb}
\ee
where $\alpha_1$, $\alpha_2$ are two constants to be determined. 

\noindent
(iii) The expanded generators $X'_k$ of $g'$  are  the elements
in the  universal enveloping algebra of $g$  defined by the
following functions of the generators $X_k$ of $g$:
\be
X'_k:=\left\{\begin{array}{ll}
X_k &\  \mbox{if}\quad [{\cal J},X_k] =0   \cr
[{\cal J},X_k] &\ \mbox{if}\quad [{\cal J},X_k]
\ne 0  
\end{array}
\right.  
\label{bbbc}
\ee

\noindent
(iv) Impose the  new generators $X'_k$ to close a Lie 
algebra isomorphic to $g'$. This gives (if possible at all) some 
conditions that characterize  the constants $\alpha_1$ and
$\alpha_2$.

Some comments are pertinent. First, this is only a proposal, not
a full-fledged method, so there is no a priori success guarantee.
Yet the method works in the cases we study,
with a {\em caveat}: in some cases the initial algebra
should be taken after being centrally extended. Second,  the  same
procedure can also be applied when  both  Lie algebras
$g$, $g'$ are higher rank. In
principle,   the operator ${\cal J}$ will have as many terms as
the number of Casimirs (the rank of the algebras).  
  In this sense, we recall that the expansion method
introduced in \cite{expansion} for the $(2+1)$D kinematical
algebras only considered Casimirs linear in the curvature. Hence
the extension of the method to higher dimensions consists
in keeping the terms which are first-order in the curvature in
order to reproduce the whole final algebra $g'$.

In the next sections we apply this method to the $(3+1)$D Galilei
algebra and  hereafter we   denote with a prime any  element 
associated to the expanded  algebra $g'$ (either Poincar\'e or
Newton--Hooke), and we  drop the prime when we  deal with the
initial one $g$ (Galilei).

%%%%%%%%%%%%%%%%% Poincar\'e %%%%%%%%%%%%%%%%%%%%%%%%%%%%%%%%%%

\sect{Recovering the speed of light: from Galilei to
Poincar\'e}

The Lie brackets of the
$(3+1)$D Poincar\'e algebra ${\cal P}\equiv iso(3,1)$ read
\be
\begin{array}{lll}
[J'_i,J'_j]=\varepsilon_{ijk}J'_k & \qquad
[J'_i,P'_j]=\varepsilon_{ijk}P'_k  &
\qquad [J'_i,K'_j]=\varepsilon_{ijk}K'_k \cr
[P'_i,P'_j]=0 &\qquad \displaystyle{[P'_i,K'_j]=-\frac{1}{c^2}
\delta_{ij}H'}     &\qquad
\displaystyle{[K'_i,K'_j]=-\frac{1}{c^2}
\varepsilon_{ijk} J'_k} \cr 
[H',P'_i]=0 &\qquad [H',K'_i]=-P'_i  &\qquad [H',J'_i]=0 . 
\end{array}
\label{ba}
\ee
 The two Poincar\'e Casimir invariants are given by
\cite{casimir}:
\bea
&&{\cal C}'_1
=\>P'^2 -\frac 1{c^2} H'^2 ={P'_1}^2+{P'_2}^2+{P'_3}^2-\frac
1{c^2} H'^2 \cr 
&&{\cal C}'_2
=  \>{W'}^2-\frac 1{c^2} (\>J'\>P')^2=
{W'_1}^2+ {W'_2}^2+{W'_3}^2 -\frac 1{c^2}
(J'_1P'_1+J'_2P'_2+J'_3P'_3)^2 
\label{bb}
\eea
where the  components of $\>{W'}$ are  
\bea
&&W'_1=-\frac 1{c^2} H' J'_1+ P'_3K'_2 -P'_2 K'_3  \cr
&&W'_2= -\frac 1{c^2} H' J'_2+  P'_1 K'_3-P'_3K'_1   \cr
&&W'_3= -\frac 1{c^2} H' J'_3+ P'_2K'_1 -P'_1 K'_2.
\label{bc} 
\eea
The Casimir $\expandcasa$ is the energy of the particle  in the
free kinematics in the Minkowskian spacetime, while 
 $\expandcasb$ is the square of the Pauli--Lubanski vector.

The Cartan decompositions  (\ref{ae}) also hold for the
Poincar\'e algebra although in this case the vector subspaces
$p^{(l)}$  satisfy $[p^{(1)},p^{(1)}]=0$ and
$[p^{(2)},p^{(2)}]\subset h^{(2)}$ in the relations
(\ref{af}). Therefore   the Poincar\'e group  $P\equiv
ISO(3,1)$ is the motion group of the symmetric homogeneous
spaces given by
\be
\begin{array}{ll}
 {\cal S}^{(1)}=P/H^{(1)}=ISO(3,1)/SO(3,1) &\qquad
\mbox{curv}\,({\cal S}^{(1)})=0\cr
{\cal S}^{(2)}=P/H^{(2)}=ISO(3,1)/(\R\otimes
SO(3)) &\qquad
\mbox{curv}\,({\cal S}^{(2)})=-1/c^2     
\end{array} 
\label{bbc}
\ee
 where ${\cal S}^{(1)}$ and ${\cal S}^{(2)}$ are identified 
with the  $(3+1)$D Minkowskian spacetime   and the 6D space of
time-like lines in Minkowskian spacetime, respectively.

The non-relativistic limit $c\to \infty$  leads to  the
contraction $\poinc\to \gal$ and makes the curvature of
${\cal S}^{(2)}$ to vanish so that the commutators 
$[P_i,K_j]$, $[K_i,K_j]$ are equal to zero in $\gal$. Our aim
now is to reverse this process, that is, to consider  $\gal$ as
the initial Lie algebra, thus  trying to recover $\poinc$ by 
introducing the relativistic constant $c$: $\gal\to \poinc$. 

According to the first step of the expansion method (\ref{bbba})
we rewrite the Poincar\'e Casimirs so as to explicitly display
the powers of the curvature
$-1/c^2$ of the space of worldlines ${\cal S}^{(2)}$: 
\be
\begin{array}{l}
\displaystyle{{\cal C}'_1={\cal C}_1  -\frac 1{c^2}  {\cal J}_1
\qquad {\cal J}_1=H^2 \qquad {\cal M}_1=0 }\cr
\displaystyle{{\cal C}'_2={\cal C}_2 -\frac 1{c^2} {\cal J}_2 +
\left(-\frac 1{c^2}\right)^2 {\cal M}_2  \qquad
{\cal J}_2=2 H\,\>J\>W + (\>J\>P)^2 \qquad {\cal
M}_2=H^2\,\>J^2  }
\end{array} 
\label{cf} 
\ee
where ${\cal C}_1$, ${\cal C}_2$ are  the Galilei Casimirs
(\ref{ab}) and  $\>W$  is defined in  (\ref{ac}).
Note that in this case, the power series in the curvature
$-1/c^2$ (\ref{bbba}) are first-order for ${\cal C}'_1$ and  
second-order for ${\cal C}'_2$. The second step of the procedure
(\ref{bbbb}) gives the linear combination
\be
{\cal J}=\alpha_1 {\cal J}_1 + \alpha_2{\cal J}_2=
\alpha_1 H^2 + 2 \alpha_2  H\,\>J\>W +\alpha_2 (\>J\>P)^2.
\label{ch}
\ee
Hence, the new generators that we want  to close the
Poincar\'e algebra  $X'_k$,  are obtained by commuting 
${\cal J}$ with the Galilei generators $X_k$ following  the third
step (\ref{bbbc});  they turn out to be
\bea
&&H'=H\qquad J'_1=J_1\qquad J'_2=J_2\qquad J'_3=J_3\cr
&&P'_1=2\AA_2 H(P_2\ww3- P_3\ww2)\cr
&&P'_2=2\AA_2 H(P_3\ww1- P_1\ww3)\cr
&&P'_3=2\AA_2 H(P_1\ww2- P_2\ww1)\label{bg}\\
&&K'_1=-2\AA_1 HP_1-2\AA_2 \>J\>W\, P_1  
+ 2 \AA_2 H(K_2 \ww3 -
K_3\ww2)\cr 
&&\qquad\quad + 3 \AA_2 (P_2 \ww3 - P_3\ww2)+ 2 \AA_2
\>J\>P\,\ww1\cr 
&&K'_2=-2\AA_1 HP_2-2\AA_2 \>J\>W\, P_2 + 2 \AA_2
H(K_3 \ww1 - K_1\ww3)\cr 
&&\qquad\quad + 3 \AA_2 (P_3 \ww1 -
P_1\ww3)+ 2 \AA_2 \>J\>P\,\ww2\cr 
&&K'_3=-2\AA_1 HP_3-2\AA_2
\>J\>W\, P_3 + 2 \AA_2 H(K_1 \ww2 - K_2\ww1)\cr 
&&\qquad\quad + 3
\AA_2 (P_1 \ww2 - P_2\ww1)+ 2 \AA_2 \>J\>P\,\ww3 .
\nonumber
\eea
Some commutators between  elements of the
universal enveloping Galilei algebra which are useful in the
obtention of the above `expanded' generators as well as in
further computations are given in the Appendix.

The last step is to impose the expanded generators (\ref{bg}) to
fulfil the Lie brackets of the Poincar\'e algebra  (\ref{ba});
this leads  to two conditions for the constants
$\alpha_1$, $\alpha_2$. The resulting expansion process  is
summarized by

\noindent
{\bf Theorem 1.} {\em The generators defined by   (\ref{bg})
close the $(3+1)$D  Poincar\'e  algebra whenever  the constants
$\AA_1$, $\AA_2$ satisfy
\be
\AA_1 \ppa + \AA_2\ppb=0\qquad
\AA_2^2 = \frac 1{4 c^2 \ppa\ppb}  
\label{bh}
\ee
where $c$ is the relativistic constant (the speed of light)  and
$\ppa$, $\ppb$ are the Galilei Casimirs (\ref{ab}) which within
an irreducible representation of $\gal$ turn into scalar
values.}

\noindent
{\em Proof.} The generators which  are invariant in this
expansion   span  the isotopy subalgebra  of a time-like line
$h^{(2)}=\langle H,\>J\rangle$.  Hence by taking into account
the second  Cartan decomposition (\ref{ae}) it can be checked
that the assumptions of the  proposition 1 of \cite{expansion}
are automatically fulfilled,  which in turn implies that the 
Lie brackets either between  two generators belonging to
$h^{(2)}$, or between one generator of  $h^{(2)}$ and another
of $p^{(2)}$ remain in the same form as in the Galilei
algebra, as it should be for  Poincar\'e. Consequently,  we
have only to compute the 15 commutators that involve two
generators of $p^{(2)}$: $K'_i$, $P'_i$.  We distinguish four
types of Lie brackets:  $[P'_i,P'_j]$,  $[P'_i,K'_j]$   which
must vanish again, and $[P'_i,K'_i]$, $[K'_i,K'_j]$   which
according to (\ref{ba})  must be now different from zero.

As it is shown  in (\ref{apendd}), $H$, $P_i$ and $W_j$
commute amongst themselves, so that the three commutators
$[P'_i,P'_j]$ are directly equal to zero. For the second type,
let us compute for instance $[P'_1,K'_2]$; if we consider
(\ref{apende}) and (\ref{apendf}) then we obtain that
\bea
&&[P'_1,K'_2]=-4 \AA_2^2 H[P_2\ww3- P_3\ww2,\>J\>W] P_2
+4 \AA_2^2 H [H, K_3 \ww1 - K_1\ww3] (P_2\ww3- P_3\ww2)\cr
&&\qquad\qquad\quad +4 \AA_2^2 H[P_2\ww3- P_3\ww2,\>J\>P]\ww2 .
\eea
We introduce (\ref{apendg}), (\ref{apendh}) and
(\ref{ab}) thus finding that  this bracket is equal  to zero
due to the identity (\ref{apenda}):
\bea
&&[P'_1,K'_2]=-4 \AA_2^2 H\left\{
P_1 P_2  \ppb +   (P_3 \ww1 - P_1\ww3) (P_2\ww3- P_3\ww2)
 +\ppa \ww1\ww2 \right\}\cr
&&\qquad\qquad 
=-4 \AA_2^2 H (P_2\ww1+P_1\ww2)(P_1\ww1+P_2\ww2+P_3\ww3)=0 .
\eea
The same
happens for the five remaining Lie  brackets $[P'_i,K'_j]$.
Similar computations allow us to deduce the three commutators
$[P'_i,K'_i]$, all of them leading to same condition.
Let us choose
\bea
&&[P'_1,K'_1]=-4 \AA_2^2 H[P_2\ww3- P_3\ww2,\>J\>W] P_1
+4 \AA_2^2 H [H, K_2 \ww3 - K_3\ww2] (P_2\ww3- P_3\ww2)\cr
&&\qquad\qquad\quad
 +4 \AA_2^2 H[P_2\ww3- P_3\ww2,\>J\>P]\ww1 \cr
&&\qquad\qquad =-4 \AA_2^2 H\left\{
P_1^2  \ppb +    (P_2\ww3- P_3\ww2)^2
 +\ppa \ww1^2 \right\} .
\eea
By   expanding, substituting the term $-2 P_2 P_3 \ww2\ww3$
from (\ref{apendc}) and imposing the corresponding
Poincar\'e bracket (\ref{ba}), we obtain that
\be
[P'_1,K'_1]=-4 \AA_2^2 H \ppa\ppb \equiv 
-\frac{1}{c^2}  H' 
\ee
which leads to  the second relation of (\ref{bh}).  

The last step is to compute the  three commutators
$[K'_i,K'_j]$; each of them gives rise to the two conditions
(\ref{bh}). Let us consider  
\bea
&&[K'_1,K'_2]=
8\AA_1\AA_2 H\left\{ P_3 (P_1\ww1 + P_2\ww2) -P_1^2\ww3
-P_2^2\ww3 \right\} -4 \AA_2^2 H \ppb (P_2 K_1 -P_1 K_2)\cr
&&\qquad\qquad
- 4 \AA_2^2 H\left\{ (P_3K_2
-P_2K_3)\ww1
+(P_1K_3-P_3K_1)\ww2+(P_2K_1-P_1K_2)\ww3\right\}\ww3\cr
&&\qquad\qquad
+ 4 \AA_2^2 \>J\>W\left\{P_1 P_3\ww1 + P_2 P_3\ww2  -P_1^2\ww3
-P_2^2\ww3 \right\}\cr 
&&\qquad\qquad 
-4 \AA_2^2 \>J\>P  \left\{ P_3 \ww1^2 +P_3 \ww2^2 -
 P_1\ww1\ww3 - P_2\ww2\ww3   \right\}\cr
 &&\qquad\qquad  
-4 \AA_2^2 [ \>J\>W, \>J\>P]  (P_1\ww2 -P_2\ww1).
\eea
In the terms that  multiply $H$, we use the identity
(\ref{apenda}) as $P_1\ww1 + P_2\ww2=- P_3\ww3$,   introduce
the components $W_i$ (\ref{ac})  and group terms   in order to
`construct' the Galilei Casimirs (\ref{ab}). Next we write the
commutator (\ref{apendi}) and group the terms multiplying each
generator of rotations $J_i$. It can be checked that those
factors associated to $J_1$ and $J_2$ vanish directly. This
gives
\bea
&&[K'_1,K'_2]=
-8 \AA_2 H \left\{\AA_1\ppa+ \AA_2\ppb \right\}\ww3
+ 4 \AA_2^2 J_3
\bigl\{
2  P_1P_2\ww1\ww2+2P_1P_3\ww1\ww3\cr
&&\qquad\quad +2P_2P_3\ww2\ww3
-P_1^2(\ww2^2+\ww3^2)
-P_2^2(\ww1^2+\ww3^2)-P_3^2(\ww1^2+\ww2^2)
\bigr\}
\eea
and by applying the  identity (\ref{apendb}) we finally obtain
that
\be
[K'_1,K'_2]=-8 \AA_2 H \left\{\AA_1\ppa
+ \AA_2\ppb \right\}\ww3
-4 \AA_2^2 J_3 \ppa\ppb \equiv -\frac 1{c^2} J'_3
\ee
so that the theorem 1 is proven.

Consequently, this Galilei Lie algebra
expansion  reverses the non-relativistic contraction of the
Poincar\'e algebra   and allows us to  introduce a constant
negative curvature $\omega=-1/c^2$ in the 6D space of worldlines
${\cal S}^{(2)}$, thus replacing the flat Galilean space of
worldlines by the (curved) space
$P/{H}^{(2)}$ (\ref{bbc}).   (Recall however that the space ${\cal
S}^{(2)}$ is rank-two, so its geometry is {\em very} different from a
rank-one 6D space of constant curvature, either in the curved or in the
flat case).   This procedure keeps  flat the
ordinary spacetime, but the change from  zero to non-zero
curvature in the space of worldlines is unavoidably linked to the
change from the Galilean `absolute time' to the Minkowskian
`relative time' nature. At the level of the Cartan
decompositions the expansion gives
$[{p}^{(2)},{p}^{(2)}]=0\to [{p}^{(2)},{p}^{(2)}]\subset 
{h}^{(2)}$.

We would like to stress that the  same procedure also holds
for the expansion going from Galilei to the  Euclidean algebra
${\cal E}\equiv iso(4)$. If we replace in {\em all} the
Poincar\'e expressions the negative curvature $-1/c^2$ by a
positive constant  then we obtain the commutation rules,
Casimirs, etc.\ corresponding to $iso(4)$; note that this is
equivalent to set $c$ equal to a pure imaginary complex
number. The expansion $\gal\to {\cal E}$  would lead to
similar `expanded' generators  and conditions  as those
characterized by the theorem 1.  In this case, the symmetric
space 
${\cal S}^{(1)}= ISO(4)/SO(4)$ is the 4D flat Euclidean
space and  ${\cal S}^{(2)}=ISO(4)/(\R\otimes
SO(3))$ is a rank-two and positively curved 6D space of 
lines in Euclidean space. 

%%%%%%%%%%%%%%%%% Newton--Hooke %%%%%%%%%%%%%%%%%%%%%%%

\sect{Recovering the universe time radius: from extended
Galilei to Newton--Hooke}

Besides the Poincar\'e and Euclidean algebras, the
Galilei algebra has other two physically  remarkable neighbours:
the oscillating (or anti) and the expanding Newton--Hooke algebras
\cite{BLL}, hereafter denoted ${\cal N}_+$ and ${\cal N}_-$,
respectively.  Both of them are the Lie algebras of  the
motion groups of absolute time universes but with non-zero
curvature $\kappa$.  The commutation rules of  ${\cal N}_\pm$
are given  by 
\be
\begin{array}{lll}
[J'_i,J'_j]=\varepsilon_{ijk}J'_k & \qquad
[J'_i,P'_j]=\varepsilon_{ijk}P'_k  &
\qquad [J'_i,K'_j]=\varepsilon_{ijk}K'_k \cr
[P'_i,P'_j]=0 &\qquad  [P'_i,K'_j]=0    &\qquad  [K'_i,K'_j]= 0
\cr  [H',P'_i]=\kappa K'_i &\qquad [H',K'_i]=-P'_i  &\qquad
[H',J'_i]=0   
\end{array}
\label{ca}
\ee
where the curvature $\kappa$ can be expressed in terms of 
the universe time radius $\tau$ either by $\kappa=1/\tau^2$
for  ${\cal N}_+$, or by $\kappa=-1/\tau^2$ for  
${\cal N}_-$; notice that $\tau$ is a characteristic time 
\cite{BLL} so that  is measured in time units.
The Casimirs of ${\cal N}_\pm$ turn out to be 
\cite{casimir}:
\bea
&&{\cal C}'_1
=\>P'^2 +\kappa \>K'^2  ={P'_1}^2+{P'_2}^2+{P'_3}^2 
+\kappa ({K'_1}^2+{K'_2}^2+{K'_3}^2) \cr 
&&{\cal C}'_2
=  \>{W'}^2 =
{W'_1}^2+ {W'_2}^2+{W'_3}^2
\label{cb}
\eea
where the components of $\>{W'}$ are formally identical to the
Galilei ones (\ref{ac}).

Due to the non-zero Lie brackets $[H',P'_i]=\kappa K'_i$,   the
vector subspaces $p^{(l)}$ of the Cartan decompositions  
(\ref{ae})  verify now $[p^{(1)},p^{(1)}]\subset
h^{(1)}$ and $[p^{(2)},p^{(2)}]=0$. The Newton--Hooke groups 
${N}_\pm$ are the motion groups  of the following  symmetric
homogeneous spaces 
\be
\begin{array}{ll}
 {\cal S}^{(1)}=N_+/H^{(1)}=T_6(SO(2)\otimes SO(3))/ISO(3) 
&\quad \mbox{curv}\,({\cal S}^{(1)})=1/\tau^2\cr
{\cal S}^{(2)}=N_+/H^{(2)}=T_6(SO(2)\otimes
SO(3))/(SO(2)\otimes SO(3)) &\quad
\mbox{curv}\,({\cal S}^{(2)})=0   
\end{array} 
\label{cc}
\ee
\be
\begin{array}{ll}
 {\cal S}^{(1)}
=N_-/H^{(1)}=T_6(SO(1,1)\otimes SO(3))/ISO(3) &\quad
\mbox{curv}\,({\cal S}^{(1)})=-1/\tau^2\cr
{\cal S}^{(2)}=N_-/H^{(2)}
=T_6(SO(1,1)\otimes SO(3))/(SO(1,1)\otimes
SO(3)) &\quad
\mbox{curv}\,({\cal S}^{(2)})=0   
\end{array} 
\label{cd}
\ee
The spaces  ${\cal S}^{(1)}$ and  ${\cal S}^{(2)}$ 
correspond, in this order, to the   $(3+1)$D non-relativistic
curved spacetime   and the 6D flat  space of worldlines in
these spacetimes; we remark that the natural metric in this last
flat space is definite positive for ${\cal N}_+$ ($\kappa>0$) and
indefinite, with signature $(3,3)$ for ${\cal N}_-$ ($\kappa<0$).
 We recall that a recent study of the metric structure   of 
both Newton--Hooke spacetimes has been carried out in
\cite{Aldrovandi}.

The limit $\kappa\to
0$, or equivalently $\tau\to \infty$, produces the spacetime
contraction ${\cal N}_\pm\to {\cal G}$. Now we consider the
opposite situation and  as in the previous section we take
${\cal G}$ as the initial Lie algebra analysing the way of
obtaining ${\cal N}_\pm$ by introducing the universe time
radius $\tau$ (i.e.\ $\kappa$): ${\cal G}\to {\cal N}_\pm$.
We apply our expansion method  and decompose the Casimirs
(\ref{cb}) according to the curvature $\kappa$:
\be
\begin{array}{lll}
\expandcasa=\ppa  +\kappa   \jja&\qquad
\jja= \>K^2&\qquad  {\cal M}_1=0\cr
\expandcasb=\ppb &\qquad
\jjb=0   &\qquad  {\cal M}_2=0
\end{array} 
\label{cchh}
\ee
where $\ppa$, $\ppb$ are the Galilei Casimirs  (\ref{ab}).
Therefore the linear combination depending on the linear terms
in the curvature (\ref{bbbb}) is simply
\be
 \jjj= \AA_1 \jja =  \AA_1 \>K^2
\label{ci}
\ee
where $\AA_1$ is a constant  to be determined.
The relation (\ref{bbbc}) gives  the expanded generators; they
are $H'=2\AA_1 \>K\>P$ and all the remaining ones are
unchanged. It can be straightforwardly checked that these
new generators {\em do not span} the Newton--Hooke algebras; it is
necessary to take an initial Lie algebra  {\em less abelian}
in order to be able to perform the expansion. The natural choice is
to start from the (centrally) 
extended Galilei algebra. This was exactly
what was required in the
$(2+1)$D case \cite{expansion}. Hence we introduce a central
extension, with central generator $\Xi$ and parameter $\ext$
(the mass of the particle). The commutation rules of the
centrally extended Galilei algebra $\overline\gal$ are given
by (\ref{aa}) once  the vanishing bracket $[P_i,K_j]$ is
replaced by  
\be
[P_i,K_j]=\delta_{ij}\ext\Xi\qquad
[\Xi,\,\cdot\,]=0 .
\ee
We apply (\ref{bbbc}) with the element (\ref{ci})  for the Lie
brackets of $\overline\gal$,  finding that   the  new expanded
generators read
\bea
&& J'_i=J_i\qquad K'_i=K_i\qquad  i=1,2,3\cr
&&H'=2\AA_1 \>K\>P + 3 \AA_1  \ext \Xi\qquad
P'_i=-2\AA_1  \ext \Xi K_i .
\label{cj}
\eea
This expansion is characterized by

\noindent
{\bf Theorem 2.} {\em The generators defined by   (\ref{cj})
give rise to the $(3+1)$D  Newton--Hooke algebras provided
that    the constant $\AA_1$ fulfils
\be
\AA_1^2 = -\frac {\kappa}{4 \ext^2 \Xi^2}
= \mp \frac {1}{4 \tau^2 \ext^2 \Xi^2}
\qquad \mbox{for}\quad {\cal N}_\pm .
\label{ck}
\ee}

\noindent
{\em Proof.}  The generators which are unchanged  in the
expansion close the isotopy subalgebra of an event (a point in
the spacetime) $h^{(1)}= \langle \>K,\>J\rangle$. Thus the  
proposition 1 of \cite{expansion} can be applied and we only
need to compute the 6 Lie brackets involving the generators
of  $p^{(1)}$: $H'$ and
$P'_i$. By direct computations we obtain
\be
[P'_i,P'_j]=0\qquad
[H',P'_i]=-4\AA_1^2\ext^2\Xi^2 K_i\equiv \kappa K'_i
\ee
and the relation (\ref{ck}) is proven.

Therefore  this expansion introduces a constant
 curvature $\kappa$ in the flat  spacetime
$G/{H}^{(1)}$ (\ref{ag}), leading to  
curved spacetimes $N_\pm/{H}^{(1)}$ (\ref{cd}). At the level of
the Cartan decompositions this corresponds to the transition
$[{p}^{(1)},{p}^{(1)}]=0\to [{p}^{(1)},{p}^{(1)}]\subset 
{h}^{(1)}$.

%%%%%%%%%%%%%%%%% Concluding remarks %%%%%%%%%%%

\sect{Concluding remarks}

The expansion method proposed in
\cite{expansion} for the  $(2+1)$D kinematical algebras 
has been improved in order to be applied to higher dimensions and
explicitly tested with the 
$(3+1)$D Galilei algebra; the method works in the two physically
meaningful `expansion directions' which lead from Galilei spacetime
to either the flat relativistic Minkowskian spacetime or to the
curved Newton--Hooke non-relativistic spacetimes. In order to
summarize, we represent the expansions   studied in this paper
collectively in the following diagram:

\bigskip
{\footnotesize{
\begin{tabular}{ccccc}
 & &${\cal E}\equiv iso(4)$& &\\
& &$\kappa=0$, $c$ imaginary& &\\
& &4D Euclidean space& &\\[0.2cm]
& &$\uparrow$& &\\[0.2cm]
${\cal N}_+\equiv t_6(so(2)\oplus
so(3))$&$\longleftarrow$&
$\gal\equiv iiso(3)$ or 
$\overline{\gal}\equiv 
\overline{iiso(3)}$&$\longrightarrow$&$
{\cal N}_-\equiv t_6(so(1,1)\oplus
so(3))$\\ 
$\kappa=+1/\tau^2$, $c=\infty$& &
$\kappa=0$, $c=\infty$& &$\kappa=-1/\tau^2$, $c=\infty$\\
Oscillating NH & &Galilean  & 
&Expanding NH \\ 
(3+1)D spacetime& &(3+1)D spacetime& 
&(3+1)D spacetime\\[0.2cm]
 & &$\downarrow$& &\\[0.2cm]
 & &$\poinc \equiv iso(3,1)$& &\\
 & &$\kappa=0$, $c$ finite& & \\
  & &Minkowskian & &  \\
 & &(3+1)D spacetime& &  \\[0.1cm]
\end{tabular}
}} 
\bigskip

The method uses as `expansion seed' an operator built out from the
ordinary expansion in powers of a parameter ---interpreted as a
curvature--- of the Casimir operators of the expanded algebra. 
The number of these terms (or the number
of Casimirs involved in the expansion) is equal to the rank of the
homogeneous space behind the expansion. 

In the $(3+1)$D case we have discussed in detail, there is a
fourth-order Casimir  (of Pauli--Lubanski type), depending {\em
cuadratically} on the curvature. However the terms  of the expanded
Casimirs required  in the seed are only those which are  first-order in
the curvature to recover. We have done some attempts by using a `seed'
which keeps terms which are  higher order in the curvatures, but the
result seems to coincide with the one obtained by restricting the seed
to have only `first-order' terms in the curvature. Why this happens is 
an intriguing property.

The spacetime ${\cal S}^{(1)}$  is a rank-one space and  when we 
endow it with a non-zero  curvature  obtaining  the Newton--Hooke
algebras, the linear combination  $\jjj$ has a single term
(\ref{ci}). On the other hand, the space of time-like lines ${\cal
S}^{(2)}$ has  rank-two and   when we introduce the curvature only
in the space of time-like worldlines,  reaching the Poincar\'e
algebra, the combination $\jjj$ has two  terms  (\ref{ch}) so that
in this case both Casimirs are essential. In this sense, we remark
that all other explicit  procedures already known for expanding 
Poincar\'e or Euclidean algebras to the simple pseudo-orthogonal
algebras (\ref{intro})  are rank-one (they
introduce curvature in ${\cal S}^{(1)}$) and involve only the
quadratic Casimir \cite{Gilmore}. These can also be  obtained  (in any
dimension) by following our  expansion method,  thus recovering the
results given in  \cite{Gilmore}.  

Finally, it is worth mentioning that we have only  worked at a
Lie algebra level. Hence an interesting open problem which
naturally arises is to analyze how to implement this kind of
processes in the   representation theory. Recall that 
contractions of representations have
 been already formulated (see, e.g., \cite{MooPat}).

%%%%%%%%%%%%%%%%% ACKNOWLEDGMENTS %%%%%%%%%%%

\noindent {\section*{Acknowledgments}}

\noindent
This work was partially supported by DGES   (Project
PB98--0370) from the Ministerio de Educaci\'on y Cultura  de
Espa\~na and by Junta de Castilla y Le\'on (Project  CO2/399).

\bigskip

%%%%%%%%%%%%%%%%% APPENDIX %%%%%%%%%%%

%\newpage

\noindent
{\section*{Appendix: some relations in the universal  
enveloping Galilei algebra}}

\appendix

\setcounter{equation}{0}

\renewcommand{\theequation}{A.\arabic{equation}}

\noindent
We present some useful Lie brackets between elements of the
universal enveloping Galilei algebra which are needed   in the
proof of the expansion to the Poincar\'e algebra.  First, we
write down some {\em identities} which are  used  in
the obtention of the commutators listed below as well as in
the computations of the expansion:
\be
\>P \>W = P_1 \ww1 + P_2\ww2 + P_3\ww3=0\qquad
\>K \>W = K_1 \ww1 + K_2\ww2 + K_3\ww3=0 .
\label{apenda}
\ee
The first identity leads to other four ones:
 \be
-2P_1P_2\ww1\ww2-2P_1P_3\ww1\ww3-2P_2P_3\ww2\ww3=
P_1^2\ww1^2+P_2^2\ww2^2+P_3^2\ww3^2
\label{apendb}
\ee
\be
\begin{array}{l}
P_1^2\ww1^2-P_2^2\ww2^2-P_3^2\ww3^2-2P_2P_3\ww2\ww3=0\cr
P_2^2\ww2^2-P_1^2\ww1^2-P_3^2\ww3^2-2P_1P_3\ww1\ww3=0\cr
P_3^2\ww3^2-P_1^2\ww1^2-P_2^2\ww2^2-2P_1P_2\ww1\ww2=0 .
\end{array}
\label{apendc}
\ee 
Likewise, the second relation in (\ref{apenda}) can be   used
to obtain four   identities analogous to (\ref{apendb}) and
(\ref{apendc}) but with the generators $K_i$ instead  of the
$P_i$.

Now we display the Lie brackets between  the Galilei
generators,  the components $\ww{i}$ and the products
$\>J\>P$, $\>J\>W$:
\be
\begin{array}{l}
[\ww{i},H]=0\qquad [\ww{i},J_j]=\varepsilon_{ijk}\ww{k}\qquad
[\ww{i},P_j]=0\qquad [\ww{i},K_j]=0\qquad 
[\ww{i},\ww{j}]=0\end{array}
\label{apendd}
\ee 
\be
\begin{array}{l}
[\>J\>P,H]=0\qquad [\>J\>P,J_i]=0\qquad
[\>J\>P,P_i]=0\qquad [\>J\>P,K_i]=\ww{i}\cr
[\>J\>P,\ww{1}]=-(P_2\ww3 - P_3\ww2)\cr
[\>J\>P,\ww{2}]=-(P_3\ww1 - P_1\ww3)\cr
[\>J\>P,\ww{3}]=-(P_1\ww2 - P_2\ww1)\end{array}
\label{apende}
\ee 
\be
\begin{array}{l}
[\>J\>W,H]=0\qquad [\>J\>W,J_i]=0\qquad
[\>J\>W,\ww{i}]=0\qquad \cr
[\>J\>W,P_1]=P_2\ww3 - P_3\ww2\qquad
[\>J\>W,K_1]=K_2\ww3 - K_3\ww2\cr
[\>J\>W,P_2]=P_3\ww1 - P_1\ww3\qquad
[\>J\>W,K_2]=K_3\ww1 - K_1\ww3\cr
[\>J\>W,P_3]=P_1\ww2 - P_2\ww1 
\qquad[\>J\>W,K_3]=K_1\ww2 - K_2\ww1 .
\end{array}
\label{apendf}
\ee 

Finally, other necessary  commutators are given by:
\be
\begin{array}{l}
[\>J\>P,P_2\ww3 - P_3\ww2]=\ppa\ww1 \qquad
[\>J\>P,K_2\ww3 - K_3\ww2]=\>K\>P\, \ww1\cr
[\>J\>P,P_3\ww1 - P_1\ww3]=\ppa\ww2\qquad
[\>J\>P,K_3\ww1 - K_1\ww3]=\>K\>P\, \ww2\cr
[\>J\>P,P_1\ww2 - P_2\ww1 ]=\ppa\ww3
\qquad[\>J\>P,K_1\ww2 - K_2\ww1]=\>K\>P\,\ww3
\end{array}
\label{apendg}
\ee 
\be
\begin{array}{l}
[\>J\>W,P_2\ww3 - P_3\ww2]=-\ppb P_1 \qquad
[\>J\>W,K_2\ww3 - K_3\ww2]=-\ppb K_1 \cr
[\>J\>W,P_3\ww1 - P_1\ww3]=-\ppb P_2\qquad
[\>J\>W,K_3\ww1 - K_1\ww3]=-\ppb K_2 \cr
[\>J\>W,P_1\ww2 - P_2\ww1 ]=-\ppb P_3
\qquad[\>J\>W,K_1\ww2 - K_2\ww1]=-\ppb K_3
\end{array}
\label{apendh}
\ee 
\be
[\>J\>W,\>J\>P]=J_1(P_2\ww3 - P_3\ww2) + 
J_2 (P_3 \ww1 - P_1\ww3) + J_3
(P_1\ww2 - P_2 \ww1) .
\label{apendi}
\ee

\bigskip

%\newpage

%%%%%%%%%%%%%%%%% BIBLIOGRAPHY %%%%%%%%%%%

\end{document}